**Setting up and modelling of overflowing fed-batch cultures of *Bacillus subtilis* for the production and continuous removal of lipopeptides**

J.S. GUEZ[a]*, S. CHENIKHER[b], J.Ph. CASSAR[b], P. JACQUES[a]

[a] ProBioGEM Laboratoire des Procédés Biologiques Génie Enzymatique et Microbien, UPRES-EA 1026, Polytech-Lille, Université des Sciences et Technologies de Lille, Bd Paul Langevin, 59655 Villeneuve d'Ascq, France

[b] LAGIS Laboratoire d'Automatique Génie Informatique et Signal, Polytech-Lille, UMR 8146 CNRS, Université des Sciences et Technologies de Lille, Bd Paul Langevin, 59655 Villeneuve d'Ascq, France

(*) Author for correspondence: jean-sebastien.guez@polytech-lille.fr

Tel : 0033 (0)3 28 76 74 09

Fax : 0033 (0)3 28 76 74 01

## 1. Introduction

*Bacillus subtilis* produces bioactive lipopeptides such as iturin, surfactin (Bonmatin et al., 2003) and fengycin (Schneider et al., 1999) with high potential in biotechnological and pharmaceutical applications. A lot of interest has focussed on their surface-active properties since they are recently considered as an alternative to chemical surfactants. In addition, molecules like mycosubtilin, a member of iturin family, show a high potential as an antifungal (Maget-Dana and Peypoux, 1994). A main difficulty during the production process of lipopeptide biosurfactants in aerated and stirred bioreactors lies in the high foaming properties of these surface-active compounds. They cause the massive overflow, out of the bioreactor, of the culture broth containing cells. The use of chemical antifoams is limited because of their influence on the oxygen transfer rate and their effect on the biomass physiology (Lee and Kim, 2004). The foam overflow was used by different authors to ensure the recovery of surfactin from *Bacillus subtilis* culture broth in a collector (Cooper et al., 1981, Davis et al., 2001).

In this work, a process was developed to extract lipopeptide biosurfactants like mycosubtilin and surfactin using the foaming properties of these products. Instead of trying to avoid the foaming phenomenon, the broth foaming capacity was increased by using a loaded surface impeller and the foam was collected in continuous in a collector. A feeding strategy was applied in reference to the carbon-limited exponential fed-batch culture (EFBC) protocol with a given feed rate as reference parameter. In the range of the low feed flow rates that were used, the flow rates of the overflowing foam and of the feed were equal and the broth volume was then kept constant in the bioreactor. This process is called, in the rest of the paper, overflowing exponentially fed-batch culture (O-EFBC).

Since the process operates with a constant volume, it could be related to continuous cultures.

For these cultures, the biomass concentration in effluent flow is equal to biomass concentration in the bioreactor. Analysis of biomass concentration in the foam collector clearly shows that it is not the case for the O-EFBC. Only a part of the biomass is extracted from the foam following a linear relationship. Whatever the relationship between biomass concentration in foam and in bioreactor, since the former is lower than the latter, the O-EFBC benefits from the advantage of the increase of the biomass concentration in the bioreactor. This process could be thus considered as a recycling one (McIntyre et al., 1999; Shuler et al., 2002). Due to the biomass recycling, the specific growth rate is no more constant even with exponential feed flow rate. A dynamical model is needed to simulate the evolution of the specific growth rate during the experiments.

Section 2 of the article provides the description of experimental conditions and of data collection. Section 3 addresses the modelling framework and theoretical aspects for identification of the parameters of the model. While the optimisation strategy is given in section 4, section 5 provides a discussion of experimental results and of the design of the model. It exhibits the relationship between the mean specific growth rates estimated by the model and the mycosubtilin productivity.

## 2. Materials and Methods

### *2.1. Bacterial strain and growth conditions in flask and bioreactor*

The strain BBG100 used in this study is a mycosubtilin overproducer derivative obtained by Leclère et al. (2005) from *B. subtilis* ATCC 6633 by replacement of the native promoter of the mycosubtilin synthetase operon by a promoter originating from the replication gene repU of *Staphylococcus aureus* plasmid pUB110. The inoculum was prepared in two stages. Growth medium was first inoculated by adding a loopful of cells conserved at –80°C in a rich medium containing 40 % glycerol. Cells were grown at 140 rpm and 30°C in Erlenmeyer flasks containing medium E (Clark et al., 1981). The culture was then inoculated in Erlenmeyer flasks containing the Landy medium modified as follows: Glucose, 20 g $L^{-1}$; $(NH_4)_2SO_4$, 2.3 g $L^{-1}$; Yeast extract, 1 g $L^{-1}$; Glutamic acid, 2 g $L^{-1}$; $K_2HPO_4$, 1.0 g $L^{-1}$; $MgSO_4$, 0.5 g $L^{-1}$; KCl, 0.5 g $L^{-1}$; $CuSO_4$, 1.6 mg $L^{-1}$; $Fe_2(SO_4)_3$, 1.2 mg $L^{-1}$; $MnSO_4$, 0.4 mg $L^{-1}$.

A 5 L Bioflow 3000 bioreactor (New Brunswick, NJ, USA) was inoculated with mid-logarithmic grown cells, previously washed with one volume of a sterile solution of NaCl 9 g $L^{-1}$ after a 5 min and 3000 g centrifugation. The pH was controlled at the value of 6.5 with adding KOH or $H_2SO_4$ solutions. Temperature was controlled at 30°C. The aeration rate was fixed at 0.25 VVM and the dissolved oxygen concentration was controlled above 10% of the saturation concentration thanks to the adaptive stirrer speed ranging from 200 to 400 rpm. The feed medium was a two fold concentrated Landy modified medium.

The used impellers were two Rushton turbines arranged in a specific way. The lower one was immersed in the broth, as it is done classically in a culture process, to ensure the stirring and the oxygen transfer. The upper impeller was unusually fixed just above the initial height of the liquid broth. During the fed-batch phase, this impeller was progressively immersed in the liquid broth favouring the mixing at the gas-liquid interface. The obvious effect of this loaded surface impeller was to increase drastically the foaming capacity at the gas liquid interface. The off-gas cooling devices were oversized in order to get a more compact foam in the bioreactor outgoing tubes. The foam was collected in cooled collecting devices. The software used for controlling the process and acquiring data was AFS Biocommand (New Brunswick, NJ, USA).

*2.2. The overflowing exponential fed-batch culture (O-EFBC)*

The O-EFBC process was divided in three main stages, figure 1. In phase *I*, cells were grown 24 h in a preliminary batch culture. The stirrer speed increased progressively to maintain the dissolved oxygen above 10% of the saturation concentration until the main carbon source was depleted. It is to noticed that the re-consumption of organic acid by-products like acetate was observed. In phase *I*, the volume *V* of broth in the bioreactor decreased because of the foam overflow appearing in the early stage according to the constitutive production of mycosubtilin. In phase *II*, the cells were grown in an exponential fed-batch culture where the volume *V* increased according to the exponentially feed flow rate. $f_{in}(t)$ was calculated according to the following formula derived from carbon-limited EFBC:

$$f_{in}(t) = \frac{V_0 X_0}{Y_{X/S}^{Feed}(S_{in} - S_0)} \cdot \frac{e^{t/\tau}}{\tau} = R(t_0) \cdot \frac{e^{t/\tau}}{\tau} \qquad (1)$$

Where $R(t_0)$ is the initial value of the feed flow rate calculated from the initial biomass quantity $V_0X_0$, $Y_{X/S}^{Feed}$ biomass yield on glucose used to calculate the feed rate, 0.30 g $g^{-1}$, close to the value of 0.31 g $g^{-1}$ (Lee et al., 1997), $S_{in}$ and $S_0$ the glucose concentrations in the feed medium, 40 g $L^{-1}$, and in the bioreactor at the feed beginning, close to zero gram per litre and $1/\tau$ the expected growth rate under the assumption of a pseudo steady-state EFBC. Finally, a constant volume *V* operation of the exponential fed-batch culture was achieved in phase *III* as the foam overflow rate became equal to the feed flow rate. During phases *II* and *III*, the stirrer speed increased progressively reflecting the oxygen demand of the growing cells. The lengths of these phases depend on the values of $R(t_0)$ and the feed rate $1/\tau$. The constant volume *V* operation could not be ensured for high feed rates because $f_{in}$ became higher than $f_{out}$. Eight O-EFBC experiments operated at different feed rates $1/\tau$ ranging from 0.008 to 0.086 $h^{-1}$ were done.

*2.3. Analyses*

Lipopeptides: culture and overflowed foam samples were centrifuged at 10,000g for 10 min. A volume of 1 mL of the supernatants was purified through C18 Maxi-Clean cartridges (Alltech, Deerfield, U.S.A.). The charged column was washed with 8 mL of water and 8 mL of a 50/50 water and methanol solution. The lipopeptides were then eluted with 8 mL of 100% methanol (HPLC grade, Acros Organics, Geel, Belgium). The extract was brought to dryness before dissolution in 200 µL methanol. The sample is then injected and analyzed by high-performance liquid chromatography using a C18 column (5µm, 250 x 4.6 mm, 218 TP, VYDAC). The mycosubtilins were separated with a acetonitrile/water/trifluoroacetic acid solvent, 40:60:0.5, V/V/V, and surfactins with an acetonitrile/water/TFA 80:20:0.5, V/V/V. The flow rate was 1 mL $min^{-1}$, detection wavelength is 214 nm. Purified mycosubtilins and surfactins used as standards were from Sigma. The retention time and second derivative of the absorption spectrum between 200 and 400nm (Diode Array PDA 996, Waters) were used to identify the eluted molecules (Millenium Software, Waters).

Biomass: the bacterial dry weight was determined after drying 48h at 110°C a washed pellet of a 10 mL sample. The optical density at 600 nm was read with a spectrophotometer (uV Mini 1 240, Shimadzu, Japan). An optical density of 1.0 corresponded to a biomass concentration of 0.33g D.W;$L^{-1}$.

## 3. Modelling and theoretical aspects

The model of the process must handle both EFBC and O-EFBC behaviours that were respectively observed in the phase *II* and *III* of the experiments. A dynamic model based on biomass balance in the bioreactor and in the collector is derived from the general state space dynamical model described by Bastin and Dochain, 1990.

*3.1. General dynamic model equations*

The investigated model is nonlinear in terms of both its parameters and variables. In the rest of the presentation, all the variables are functions of time. For simplicity, this mention is avoided when it is not necessary. The mass balance of biomass and substrate in the bioreactor are given by:

$$\frac{dVX}{dt} = \mu VX - X_{out} f_{out} \qquad \mu = \frac{\mu_{\max} S}{K_S + S} \tag{2}$$

$$\frac{dVS}{dt} = -(\frac{\mu}{Y_{X/S}} + m_s)VX + S_{in} f_{in} - S f_{out} \tag{3}$$

$$\frac{dV}{dt} = f_{in} - f_{out} \tag{4}$$

Where the specific growth rate is modelled in equation (2) as a Monod's function. The mass balance in the foam collector is given by:

$$\frac{dV_{col} X_{col}}{dt} = X_{out} f_{out} \tag{5}$$

In this part, the model involves 4 parameters (see Appendix A for the definitions): $\mu_{max}$, $K_S$, $m_S$ and $Y_{X/S}$ related to physiological characteristics of the microorganism. This model allows tackling phases *II* and *III* of the culture process.

*3.2. Relationship between biomass concentration in the bioreactor and in the foam*

In continuous bioprocesses without recycling, biomass concentration in the effluent flow is equal to the concentration in bioreactor $X$. However, for foam-overflowing processes, as only a part of biomass is extracted, the biomass concentration in the dragged foam $X_{out}$ can be expressed by the linear relation:

$$X_{out} = \alpha X + \beta \qquad (6)$$

Where $\alpha$ and $\beta$ coefficients are two additional parameters of the model. This relation allows handling two main assumptions. $X_{out}$ is proportional to $X$ ($\beta = 0$) that corresponds to a partial recycling process, or $X_{out}$ is independent from $X$ ($\alpha = 0$).

*3.3. Description of the foam overflowing flow rate*

In phase *II*, since the foam overflowing flow rate is very small, the output flow is considered to be zeroed, that leads to the condition: $f_{out} = 0$ L h$^{-1}$. The broth volume in bioreactor, $V$, is a function of time and depends on the feeding flow rate $f_{in}$.

In phase *III*, the process operates at a constant volume. Therefore, the foam overflowing flow rate must be equal to the feed flow rate that leads to the condition:

$$f_{out} = f_{in} \qquad (7)$$

In order to verify these two conditions and to ensure the transition between both phases, the foam overflowing flow rate can be globally calculated as a function of $V$: $f_{out} = f_{in} \dfrac{C}{V_{\lim} - V}$ where $C = 10^{-3}$ L, is a constant related to the very small overflowing of foam that remains in phase *II*. This form also guarantees the volume $V$ in the bioreactor to be constant in phase *III* and equal to $V_{\lim} - C$.

## 4. Parameter estimation strategies

### *4.1. Optimisation scheme*

Parameter estimation consists in finding a set of appropriate values for the model parameters using the available experimental data. Parameter identification of nonlinear dynamical models can be applied to biological culture process (Gao et al., 2005). A chosen objective function, subject to the constraints represented by the dynamical model is optimised according to the parameters values. The objective function reflects the discrepancies between the experimental data and the outputs of the model. It can be considered as a measurement of the quality of the fitting operation between these two datasets. A typical objective function is defined as the sum of the squared errors between the experimental dataset and the model outputs, simulated for given trial parameters values.

$$J(\theta) = \sum_{j=1}^{N_E} \sum_{i=1}^{N_S(j)} (\xi_{ij} - \hat{\xi}_{ij})^T (\xi_{ij} - \hat{\xi}_{ij}) \tag{8}$$

$$\theta^* = Arg \min_{\theta} \left( J(\theta) \right)$$

Where $N_E$ is the number of experiments, $N_S(j)$ is the number of samples in experiment $j$, $\xi$ and $\hat{\xi}$ are respectively the vectors of the measured and the model output values. $\theta$ and $\theta^*$ are respectively the vector and the optimised vector of parameters.

The nonlinear scheme of the model imposes an iterative procedure for the estimation of its parameters and the small number of available data imposes the reduction of the number of these parameters. The quality of the fit depends on the quality of the data that is affected by uncertainties inherent in the measurement and analysing processes. A good choice of initial guesses for the parameter values and boundaries is very important to reach the global minimum. The optimization procedure whose scheme is given in figure 2 makes iterative adjustments of the parameter values to drive the objective function to its minimum. Once, this objective is reached, the obtained parameter values are kept to be used in the model.

Unconstrained methods for parameter estimation do not introduce information about parameter boundaries (Syed and Phillips, 2000). In the alternate case, constrains methods are subjected to parameter boundary values (Roll et al., 2005) thus reducing the size of the domain in which the optimal solution is searched. Here, boundary values are tested between 0.19 and 0.45 g $g^{-1}$ for the growth yield on substrate and between 0.30 and 0.40 $h^{-1}$ for the maximal specific growth rate. For our application, the quadratic objective function is calculated from three experimental datasets. We choose a gradient-based method, which is implemented in Matlab™ software by fmincon function, to solve constrained nonlinear optimization problem. In this function, an estimate of the Hessian of the Lagrangian function is updated at each iteration. Simulink™ achieves the simulation procedure.

*4.2. Biomass concentration estimation in the foam overflow*

Since the amount of biomass in the foam collector is measured by the dry weight technique which requires a minimum of collected volume, the biomass concentration in the foam $X_{out}$ is not directly available. Under the assumption, that $X_{out}$ remains constant on a short enough time period $[t_1, t_2]$, $X_{out}$ can be estimated as the relation between variations of biomass quantity and volume variations in the foam collector. The concentration of the biomass in the foam overflow is estimated as a ratio of the variation of biomass quantity and the variations of volume during the time period between 2 measurements.

$$\hat{X}_{out}[t_1, t_2] = \frac{\Delta(V_{col} X_{col})_{[t_1, t_2]}}{\Delta(V_{col})_{[t_1, t_2]}} \tag{9}$$

Where $X_{col}$ is the biomass concentration in the collector and $V_{col}$ is the collector volume.

The value of the biomass concentration to be correlated with $\hat{X}_{out}[t_1, t_2]$ is the mean value of the $X$ on the time period $[t_1, t_2]$.

*4.3. Optimisation strategy*

In our experiments, the small number of samples for each experiment might not allow the proper estimation of all parameters of the model which are $\mu_{max}$, $K_S$, $m_S$, $Y_{X/S}$, $\alpha$ and $\beta$.. Therefore, a subset of pertinent parameters that can be reliably estimated from the available data has to be selected.

As the process operates at carbon limited-growth at low growth rates and according to the shape of sensitivity functions, parameters $K_S$ and $\mu_{max}$ are correlated (Noykova et al., 2000; Patnaik 1999). Estimating them simultaneously is not recommended. Therefore, $K_S$ was fixed at 0.015 g $L^{-1}$ close to the valuse of 0.010 g $L^{-1}$ (Martinez et al., 1998) and of 0.018 g $L^{-1}$ determined previously in our laboratory. The boundary values of ms were defined according to previous published values from 0.065 (Monroy et al., 1996) to 0.12 g g-1 h-1 (Sauer et al., 1996). Thus, the set of parameters to be estimated was reduced to the subset $K_S$, $Y_{X/S}$, $\alpha$ and $\beta$.

## 5. Results and discussion

*5.1. Lipopeptides extraction performances*

A process for continuous production of mycosubtilin with *Bacillus subtilis* BBG100 strain was set up based on an O-EFBC. Three experiments were performed at different feed rates $1/\tau$.. The final concentration of lipopeptides in the bioreactor and in the foam collector are given in table 1. Both, mycosubtillin and surfactin were found as expected but exclusively in the foam collected outside the bioreactor. The biosurfactant recovery percentage was thus about 100%. Similar recovery into the foam was previously described for surfactin (Davis et al., 2001) but never for any iturinic compounds.

*5.2. Estimation of the biomass concentration in the foam*

The estimation of the biomass concentration in the foam is expressed as a function of the mean concentration in the bioreactor in figure 3. It exhibits a pseudo-constant $X_{out}$ zone within a $X$ range comprised between 4 and 10 g $L^{-1}$. A linear regression performed in this zone gives $\alpha = 0.02$ and $\beta = 2.32$ g $L^{-1}$. The low value of $\alpha$ indicates that the biomass concentration in the foam $X_{out}$ is quite independent from the biomass concentration in the bioreactor $X$.

*5.3. Estimation of the model parameters*

The optimisation strategy permits the estimation of the parameter subset $\mu_{max}$, $Y_{X/S}$, $\alpha$ and $\beta$. Results are given in table 2 for both $m_S$ boundary values. With a $m_S$ of 0.12 g $g^{-1}$ $h^{-1}$, the value of $\beta$ obtained is equal to 1.70 g $L^{-1}$. As $m_S$ becomes equal to 0,065 g, the value of $\beta$ reaches 2.01 g $L^{-1}$ $g^{-1}$ $h^{-1}$ getting more close to the value of 2.32 g $L^{-1}$ determined experimentally in figure 3. The estimated value of $\alpha$ is equal to zero and thus confirms the independency of the biomass concentration in the foam $X_{out}$ from the biomass concentration in the bioreactor $X$. The value of $\mu_{max}$ of 0.33 $h^{-1}$ is also very close to the value of 0.35 $h^{-1}$ calculated during phase *I* of the process. Consequently, a $m_S$ value of 0,065 g $g^{-1}$ $h^{-}1$ was chosen in the sequel of the article for further simulations. The $Y_{X/S}$ value is equal to 0.20 g $g^{-1}$ which is slightly lower than expected according to the bibliography. In case of *Bacillus subtilis* growths on glucose, authors generally report values ranging from 0.31 (Lee et al., 1997) to 0.46 g $g^{-1}$ (Sauer et al., 1996). The low value of the growth yield on glucose is synonymous with a low energy efficiency. In our case, this phenomenon could be explained by two factors. First, acetate has been detected in the culture medium and thus represents an energetic loss. Second, the presence of a substrate like glutamic acid in the culture medium leads to the non-oxydative conversion of pyruvic acid to acetoin (Keynan et al.,1954) representing a potential supplementary loss of energy. Previous experimental measurements done in our laboratory during batch culture of *Bacillus subtilis* ATCC 6633 in shaken flasks in the presence of both glucose and glutamic acid (Guez et al., 2004) allowed the calculation of $Y_{X/S}$ values above 0.22 g $g^{-1}$ close to the value of 0.20 g $g^{-1}$ estimated by the model.

Measured and estimated biomass quantities in bioreactor and in the foam collector are represented in figure 4 (left). They correspond to the application of the optimisation strategy with $m_S$ = 0.065 g $g^{-1}$ $h^{-1}$ and $K_S$ = 0.015 g $L^{-1}$. The standard deviation of the residual errors between experimental data and the model outputs is equal to 0.83 g. It should be noticed that the error was 23% higher in the case of results simulated with a parameters subsets estimated with a $m_S$ of 0.12 g $g^{-1}$ $h^{-1}$ confirming our choice of a $m_S$ of 0.065 g $g^{-1}$ $h^{-1}$. The evolution of the measured volume in the bioreactor is represented in figure 4 (right) and fits with the estimated values.

*5.4. Correlation between mean specific growth rate and productivity*

The growth rate simulation is done and the evolution curves as functions of time for three different feed rates are given in figure 5.

For fed-batch phase *II* where foaming is negligible, the specific growth rate increases to reach the expected limit value of $\mu_{lim}=1/\tau$. The beginning of the foam overflow in phase *III* induces an increase of the slope of the $\mu$ curve. In the appendix B, we demonstrate that the limit value of the specific growth rate can be reached and is greater than $1/\tau$. Indeed this limit value is equal to:

$$\mu_{\lim} = \frac{1/\tau}{1 - \dfrac{X_{out}}{Y_{X/S} S_{in}}} + \frac{m_S}{\dfrac{S_{in}}{X_{out}} - \dfrac{1}{Y_{X/S}}}$$

Results shown in figure 5 confirm that the mean values of the specific growth rates $\mu_{mean}$ are lower than $\mu_{lim}$ for each experiment. The model was extrapolated to five additional experiments led at feed rates of 0.008, 0.022, 0.040, 0.042 and 0.062 $h^{-1}$ Simulations were performed and the standard deviation of residual errors were in the range [1.09 g, 1.43 g]. The same simulations as in figure 4 were also performed and were correlated to productivity. Figure 6 (left side) shows that the mean value of specific growth rate was a linear function of the feed rate. This result confirmed the remark about the influence of the feed rate on the effects of the phenomena of phase *II* and phase *III*. Figure 6 (right side) presents the productivity as a function of the mean values of specific growth rate $\mu_{mean}$. It shows that, in the range of feed rates that was used, the productivity can be approximated by a linear relation of the mean growth rate. Therefore, the maximum mycosubtilin productivity corresponds to the maximum growth rate. The specific productivity in mycosubtilin was maximal for feed rate of 0.086$h^{-1}$. It reached the value of 1.27 mg of mycosubtilin.$g^{-1}$ of biomass.$h^{-1}$.

## 6. Conclusion and prospects

Production of biosurfactants by aerobic micro-organisms in bioreactors is frequently limited by the foaming. O-EFBC is a bioprocess in which this problem becomes an advantage because of the high recovery percentage of the biosurfactant in the foam. In our study, the continuous removal of the mycosubtilin led to a recovery percentage approaching 100%. Foaming overflow also leads to the loss of a biomass concentration determined to be quite independent of the biomass concentration in the bioreactor and measured at about 2.3 g $L^{-1}$.

This result was confirmed from an original model that permits the handling of the cases of both exponential fed-batch and overflowing exponential fed-batch. The parameter estimation permitted the simulation of the output values of the process $VX$ and $V_{col}X_{\mathrm{col}}$ with a global standard deviation of 0.83 g for the three experiments that were used for the parameter estimation. After extrapolating the model to five additional experiments, this error remained acceptable. It showed the good ability of the model to predict the process behavior at different feed rates. Based on this knowledge, constant feeding strategies allowing the growth of the microorganism at sustainable and constant specific growth rates as it is done in partial recycling process could be considered for further physiological studies requiring a steady-state. From this model, a quasi-linear relationship between productivity and the mean specific growth rate obtained from simulations was established. According to this relationship, applying higher feed rates would probably lead to a supplementary gain in productivity. This prospect could be reached by getting a more efficient foaming using another type of impeller that would hold up a constant volume of broth during the O-EFBC. However, the productivity of 1.27 mg of mycosubtilin $g^{-1}$ of biomass $h^{-1}$ obtained with the highest tested feed rate of 0.086 $h^{-1}$ was better than any other productivity determined in batch cultures with *Bacillus subtilis* BBG100 observing a six fold improvement.

**Acknowledgements**

This work received the financial support from the Université des Sciences et Technologies de Lille, the Région Nord Pas de Calais and the European Funds for the Regional Development.

## *Appendix A*

## *List of symbols*

| | |
|---|---|
| $V$ | volume of culture medium in the bioreactor (L) |
| $V_{out}$ | volume of the overflow foaming (L) |
| $V_{col}$ | volume of culture medium in the collector (L) |
| $V_0$ | volume of culture medium at the feed beginning (L) |
| $V_{lim}$ | maximal volume of culture medium imposed by the impeller position (L) |
| $f_{in}$ | feeding flow rate (L $h^{-1}$) |
| $f_{out}$ | overflowing flow rate (L $h^{-1}$) |
| $X$ | biomass concentration in the bioreactor (g $L^{-1}$) |
| $X_{out}$ | biomass concentration in the foam (g $L^{-1}$) |
| $X_{col}$ | biomass concentration in the collector (g $L^{-1}$) |
| $X_0$ | biomass concentrations in the bioreactor at the feed beginning (g $L^{-1}$) |
| $S$ | substrate concentration in the bioreactor (g $L^{-1}$) |
| $S_{in}$ | substrate concentration in the feed (g $L^{-1}$) |
| $S_0$ | substrate concentrations in the bioreactor at the feed beginning (g $L^{-1}$) |
| $\mu$ | specific growth rate ($h^{-1}$) |
| $\mu_{max}$ | maximum specific growth rate ($h^{-1}$) |
| $\tau$ | time constant of the exponential feed rate (h) |
| $K_S$ | saturation constant (g $L^{-1}$) |
| $Y_{X/S}^{Feed}$ | growth yield on substrate (g $g^{-1}$) |
| $Y_{X/S}$ | actual growth yield on substrate (g $g^{-1}$) |
| $m_S$ | maintenance coefficient (g $g^{-1}$ $h^{-1}$) |
| $\alpha$ | coefficient of the linear model of biomass concentration in the foam |
| $\beta$ | coefficient of the linear model of biomass concentration in the foam (g $L^{-1}$) |

## *Appendix B*

*Fed-batch phase II*

For fed-batch phase, from equation (2) we have :

$$\mu = \frac{\mu_{\max} S}{K_S + S} \qquad \text{(a)}$$

At time instant $t = 0$, before the fed-batch, as $S \approx 0$ then $\mu = 0$,

For a steady state of the fed-batch culture from the model (2) to (4) we obtain $\mu_{\lim} = f_{in} / V = 1/\tau$ .

*Foam-overflowing phase III*

In the foaming phase of the process, from equation (3) we have:

$$V \frac{dS}{dt} = -\frac{\mu}{Y_{X/S}} VX + \left(S_{in} - S\right) f_{in} - m_S VX \quad \text{as} \quad \frac{dV}{dt} = 0$$

Looking for the steady value of $\mu$ imposes to consider $\mu = \mu_{lim}$ and $S$ as constants,

So, $\frac{dS}{dt} = 0$ and $\frac{\mu_{\lim}}{Y_{X/S}} VX = (S_{in} - S) f_{in} - m_S VX$ and then,

$$VX = \frac{Y_{X/S}}{\mu_{\lim} + Y_{X/S} m_S} (S_{in} - S).f_{in} \qquad \text{(b)}$$

Differentiating the two members of equation (b) leads to : $\frac{dVX}{dt} = \frac{Y_{X/S}}{\mu_{\lim} + Y_{X/S} m_S} (S_{in} - S) \frac{df_{in}}{dt}$

From feed flow rate $f_{in}$ equation (1), it can be deduced that, $\frac{df_{in}}{dt} = 1/\tau . f_{in}$ , So:

$$\frac{dVX}{dt} = \frac{Y_{X/S}}{\mu_{\lim} + Y_{X/S} m_S}(S_{in} - S).1/\tau.f_{in} \qquad \text{(c)}$$

Substituting equations (b) into (2) leads to

$$\frac{dVX}{dt} = \frac{\mu_{\lim} Y_{X/S}}{(\mu_{\lim} + Y_{X/S} m_S)}(S_{in} - S) f_{in} - X_{out} f_{in} \qquad \text{(d)}$$

From (c) and (d),

$$\frac{Y_{X/S}}{\mu_{\lim} + Y_{X/S} m_S}(S_{in} - S).1/\tau.f_{in} = \frac{\mu_{\lim} Y_{X/S}}{\mu_{\lim} + Y_{X/S} m_S}(S_{in} - S) f_{in} - X_{out} f_{in} \qquad \text{(e)}$$

As the process operates on small values of substrate $S<<S_{in}$, at steady state we obtain from equation (e):

$$1/\tau\, Y_{X/S} S_{in} = \mu_{\lim} Y_{X/S} S_{in} - X_{out} \mu_{\lim} - X_{out} m_S\, Y_{X/S}$$

$$\mu_{\lim} = \frac{Y_{X/S} S_{in}.1/\tau + Y_{X/S} m_S X_{out}}{Y_{X/S} S_{in} - X_{out}} = \frac{1/\tau}{1 - \frac{X_{out}}{Y_{X/S} S_{in}}} + \frac{m_S}{\frac{S_{in}}{X_{out}} - \frac{1}{Y_{X/S}}} = 1/\tau.\frac{1 + \frac{X_{out} m_S}{1/\tau.S_{in}}}{1 - \frac{X_{out}}{Y_{X/S} S_{in}}} > 1/\tau$$

If $X_{out}$ is constant, the steady state value of $\mu$ ($\mu_{lim}$) is greater than the reference value imposed by the feed rate.

**Figure titles**

Figure 1: Evolution of the stirrer speed and of the dissolved oxygen during an O-EFBC of *Bacillus subtilis* BBG100 operated above DO=10% of the saturation constant, at 30°C, pH=6.5 and a feed rate $1/\tau$=0.06 $h^{-1}$ (A). Evolution of the volume V in the bioreactor (–●–), of the volume $V_{col}$ in the foam collector (–▲–) and of the cumulated feed volume (—) during the O-EFBC. Phase *I* and *II* are operated at non-constant volume *V* and phase *III* at constant volume *V*. Phase *I* represents the batch phase, phase *II* the exponential fed-batch phase, phase *III* the overflowing exponential fed-batch phase.

Figure 2: Model optimisation scheme.

Figure 3: Measured biomass in the bioreactor and in the collector expressed in g/L.

Figure 4: Measured and simulated values of biomass quantities in the bioreactor (Δ, solid) and in the foam collector (□, dashed), for the three experiments operated at different feed rates $1/\tau$ =0.062 (left up), $1/\tau$ =0.071 (left middle) and $1/\tau$ =0.086 (left down) expressed in $h^{-1}$. The standard deviation of residual errors was 0.83 g. Measured and simulated values of volume in the bioreactor (Δ, solid) for the three experiments operated at different feed rates $1/\tau$ =0.062 (right up), $1/\tau$ =0.071 (right middle) and $1/\tau$ =0.086 (right down) expressed in $h^{-1}$. These values were calculated according to the parameter subset $\mu_{max}$=0.33 $h^{-1}$, $Y_{X/S}$=0.20 g $g^{-1}$, $\alpha$=0 and $\beta$=2.01 g $L^{-1}$ estimated with $m_S$=0,065 g $g^{-1}$ $h^{-1}$ and $K_S$ =0.015 g $L^{-1}$.

Figure 5: Evolution of the growth rate $\mu$ (solid curve), mean value $\mu_{mean}$ (dotted curve) and feed rate $1/\tau$ (dashed curve), for the three experiments operated at different feed rates $1/\tau$ =0.062 (up), $1/\tau$ =0.071 (middle) and $1/\tau$ =0.086 (down) expressed in $h^{-1}$. These values were calculated according to the parameter subset $\mu_{max}$=0.33 $h^{-1}$, $Y_{X/S}$=0.20 g $g^{-1}$, $\alpha$=0 and $\beta$=2.01 g $L^{-1}$ estimated with $m_S$=0,065 g $g^{-1}$ $h^{-1}$ and $K_S$ =0.015 g $L^{-1}$.

Figure 6: Productivity expressed in mg of mycosubtilin $g^{-1}$ $h^{-1}$, as functions of mean growth rate $\mu_{mean}$ expressed in $h^{-1}$ (right). Mean growth rate $\mu_{mean}$ as a function of the feed rate expressed in $h^{-1}$ (left).

**Table titles**

Table 1: Concentrations of mycosubtilin and surfactin expressed in mg $L^{-1}$ in the bioreactor and in the foam collector for O-EFBC operated at different feed rates $1/\tau$ =0.062, $1/\tau$ =0.071 and $1/\tau$ =0.086 expressed in $h^{-1}$. ND : not detected.

Table 2: Model parameters estimation $\mu^*_{\max}$ , $Y^*_{X/S}$, $\alpha^*$ and $\beta^*$ determined for $K_S$ =0.015 g $L^{-1}$.

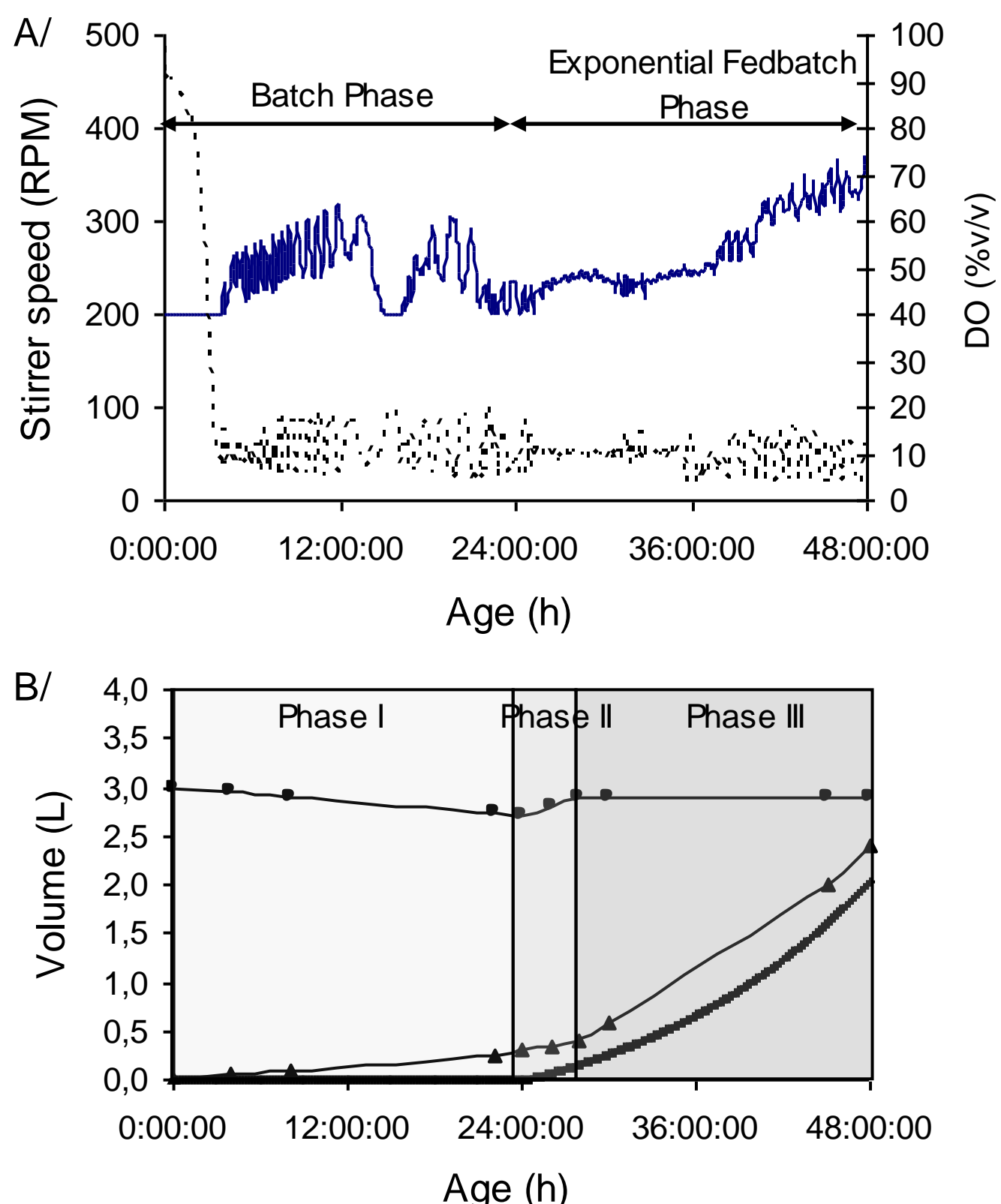

Figure 1

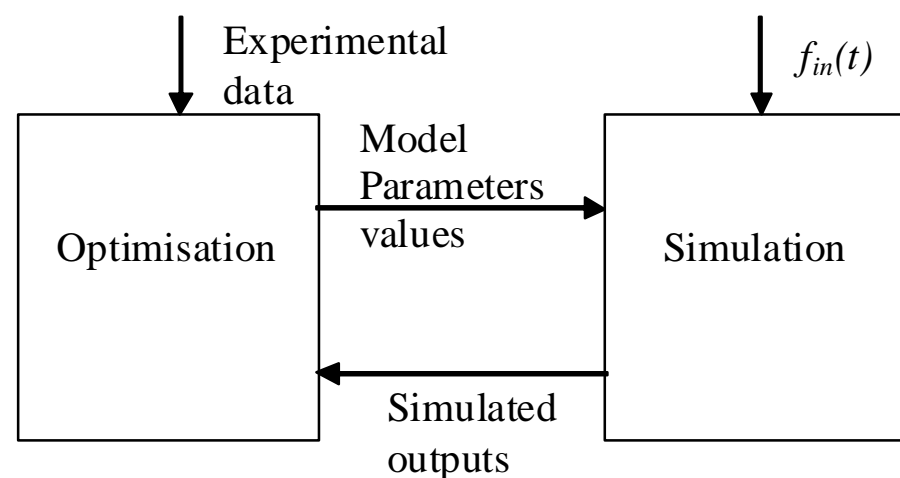

Figure 2

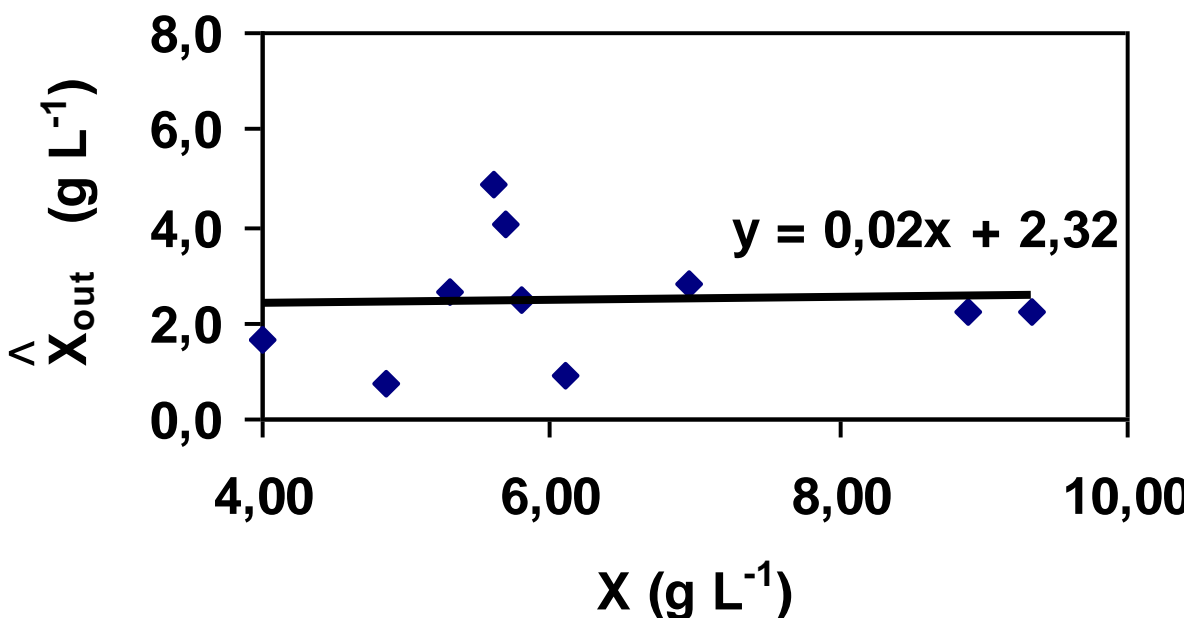

Figure 3

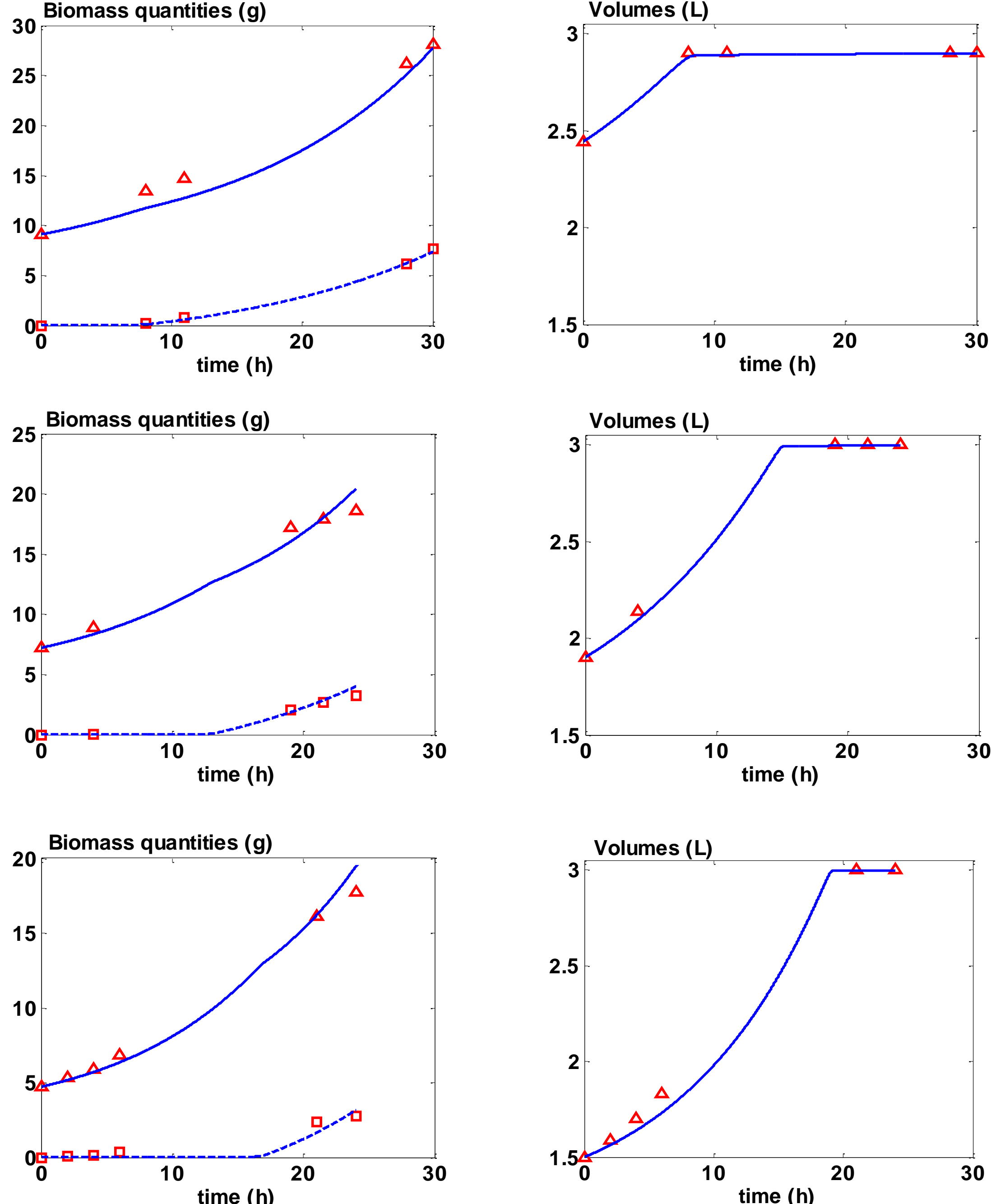

Figure 4

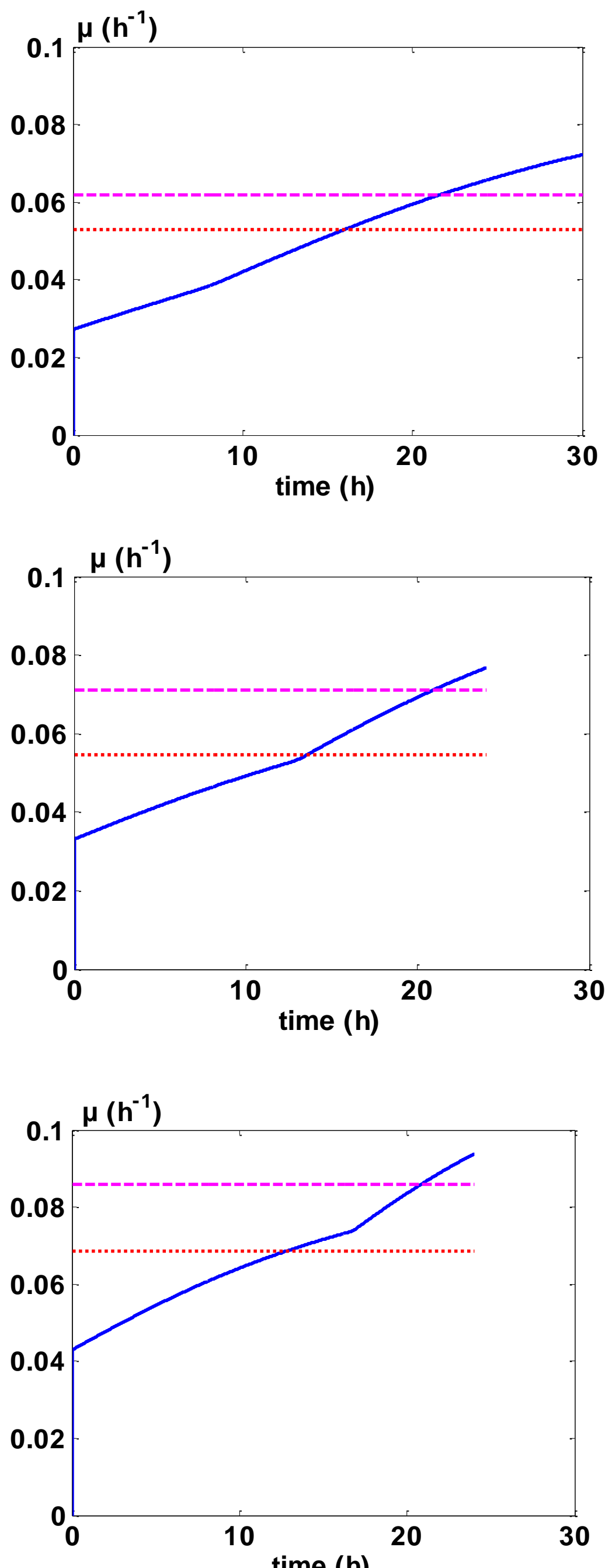

Figure 5

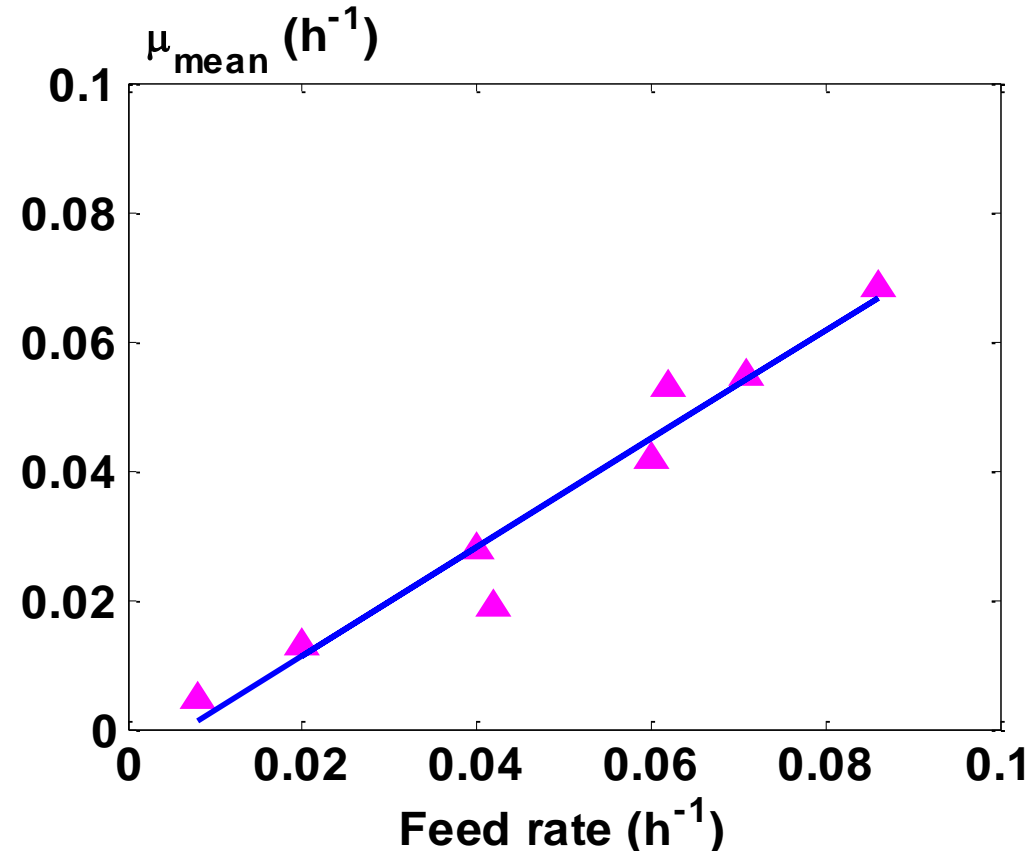

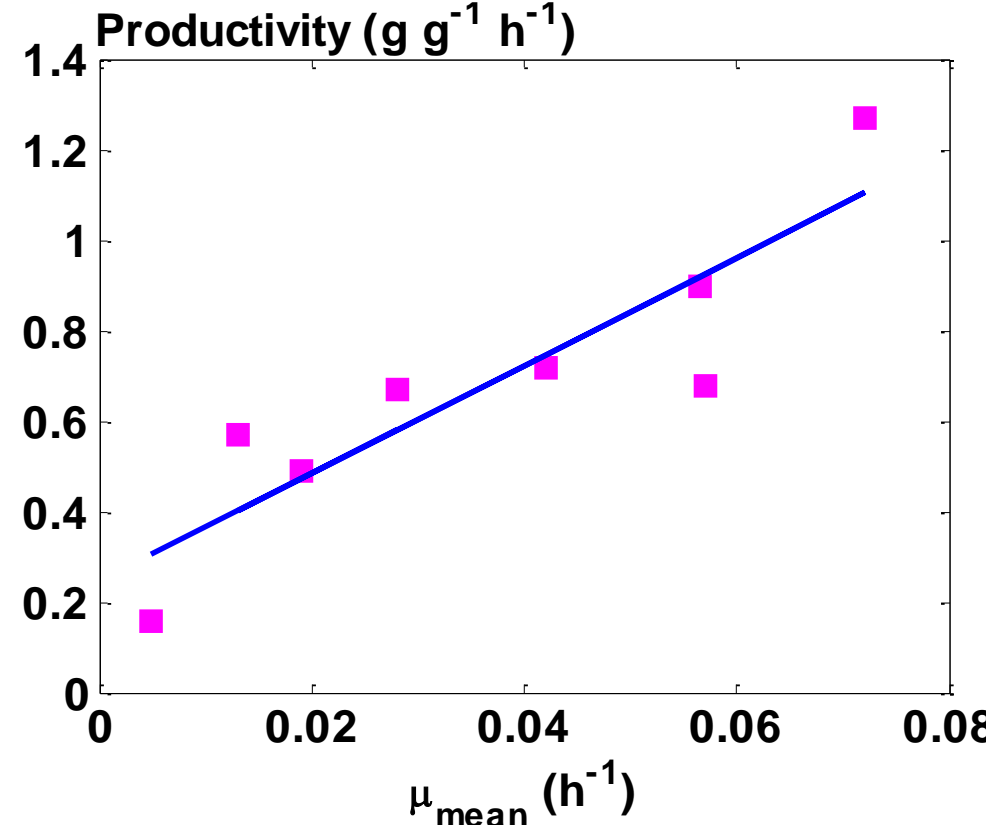

Figure 6